\documentclass[aps,pra,twocolumn,superscriptaddress,floatfix,10pt]{revtex4}
\usepackage{graphicx}
\usepackage{psfrag}
\usepackage{amssymb,amsmath}
\usepackage{physics}
\usepackage{float}
\usepackage{dsfont}
\usepackage{algorithmicx}
\usepackage[noend]{algpseudocode}
\algdef{SE}[DOWHILE]{Do}{doWhile}{\algorithmicdo}[1]{\algorithmicwhile\ #1}%

\begin{document}
\newcommand{\beq}{\begin{equation}}
\newcommand{\eeq}{\end{equation}}
\newcommand{\ben}{\begin{eqnarray}}
\newcommand{\een}{\end{eqnarray}}
\newcommand{\bea}{\begin{array}}
\newcommand{\eea}{\end{array}}
\newcommand{\om}{(\omega )}
\newcommand{\bef}{\begin{figure}}
\newcommand{\eef}{\end{figure}}
\newcommand{\leg}[1]{\caption{\protect\rm{\protect\footnotesize{#1}}}}
\newcommand{\ew}[1]{\langle{#1}\rangle}
\newcommand{\be}[1]{\mid\!{#1}\!\mid}
\newcommand{\no}{\nonumber}
\newcommand{\etal}{{\em et~al }}
\newcommand{\geff}{g_{\mbox{\it{\scriptsize{eff}}}}}
\newcommand{\da}[1]{{#1}^\dagger}
\newcommand{\cf}{{\it cf.\/}\ }
\newcommand{\ie}{{\it i.e.\/}\ }   

\newcommand{\spazio}{\vspace{0.3cm}}
\hyphenation{bio-mol-ecules}
\newcommand{\de}[1]{\frac{\partial}{\partial{#1}}}
\newcommand{\U}{\tilde{U}}
\newcommand{\V}{\tilde{V}}

\title{Optimal entangling operations between deterministic blocks of qubits encoded into single photons}

\author{Jake~A.~Smith}
\email{jsmith74@tulane.edu}
\affiliation{Tulane University, Department of Physics, New Orleans, Louisiana 70118, USA}

\author{Lev Kaplan}
\affiliation{Tulane University, Department of Physics, New Orleans, Louisiana 70118, USA}

 \begin{abstract}
Here, we numerically simulate probabilistic elementary entangling operations between rail-encoded photons for the purpose of scalable universal quantum computation or communication. We propose grouping logical qubits into single-photon blocks wherein single-qubit rotations and the CNOT gate are fully deterministic and simple to implement. Inter-block communication is then allowed through said probabilistic entangling operations. We find a promising trend in the increasing probability of successful inter-block communication as we increase the number of optical modes operated on by our elementary entangling operations.
\end{abstract}                                                               
\date{\today}
\maketitle

\section{Introduction}
 Since the discovery of several high-profile quantum algorithms in the 1990s, there has been significant interest in developing programmable quantum hardware that is both reliable and scalable. One particular approach has been the linear optical quantum computing paradigm, where qubits are encoded into the spatial and polarization states of a small number of photons. The advantage to using photons is that they tend to not interact with their environment and optical quantum states are thus naturally resistant to decoherence~\cite{Review Paper}. On the other hand, photon-photon interaction is weak and linear entangling operations are limited to bosonic interference effects~\cite{Review Paper,Hong Ou Mandel}. Early proposals for an optical quantum computer circumvented this problem by encoding information into the degrees of freedom of a single photon, but this approach limited scalability~\cite{Adami,Torma,Pittman}. In 2001, Knill, Laflamme, and Milburn discovered that universal entangling operations between a pair of qubits encoded into separate photons could be achieved probabilistically using ancillla photons and partial measurements~\cite{KLM,KLM2}. Still, experimental efforts continue to develop and use single-photon devices, often for specialized applications~\cite{Bao,Starek,Barreiro,Graham,Lanyon, Schreiber, Sansoni,Zadeh}.
 
  In our work here, we explore a balance between the reliability of single photon computing and the scalability of multi-photon computing. We propose encoding a few qubits into single-photon blocks so that within each block ``entangling'' operations between qubits are carried out through simple deterministic linear elements. Success of this model depends completely on reliable communication between blocks, which is accomplished through photon-pair entangling operations. Here, we numerically test our ability to carry out these inter-block operations using only linear optical circuit elements  (i.e., beam splitters and phase shifters), ancilla states, and probabilistic partial measurements.
  
  This paper is organized as follows: in Sections~\ref{Intro} and~\ref{Section on Protocol} we review mathematical definitions and introduce methods for simulating linear optical quantum circuits. Then, in Section~\ref{Section Block Encoding}, we discuss the single-photon block encoding model in detail. In Sections~\ref{Section Elementary Photon Entangling Operations}-\ref{Section Numerical Testing} we present a formalism for describing the most fundamental entangling operations between pairs of photons. Finally, in Sections~\ref{Section on Results}-\ref{Section Conclusion} we present our results.

\section{Mathematical Background}
\label{Intro}
An optical quantum state of $N$ photons contained in $M$ optical modes can be written in the Fock basis,
\begin{eqnarray}
	\label{LO State Fock Basis}
	\ket{\psi} = c_1 \ket{N_1,0_2,\dots,0_M} + c_2 \ket{{N-1}_1,1_2,0_3,\dots, 0_M} \nonumber \\ \dots  + c_{d_H} \ket{0_1,0_2,\dots,N_M},	 \quad \quad \quad \quad
\end{eqnarray}
where the Hilbert space dimension of the system is given by
\begin{equation}
\label{Hilbert Space Dimension}
	d_H = \frac{(N+M-1)!}{N!(M-1)!}
\end{equation}
or the number of ways to order $N$ indistinguishable photons and $M-1$ partitions.

A linear optical operation acting on a quantum state is generally described by the transformation of creation operators~\cite{Review Paper,Reck}:
\begin{equation}
\label{LO Creation Operator Transformation}
\hat{a}^\dagger_\alpha \rightarrow \sum_{\beta=1}^{M} U_{\alpha\beta} \hat{a}^\dagger_\beta \,,
\end{equation}
where $U_{\alpha \beta}$ are the elements of a unitary complex matrix, $U$. One typically writes a state given by Eq.~(\ref{LO State Fock Basis}) as
\begin{eqnarray}
\ket{\psi} = \left[c_1 \frac{(\hat{a}^\dagger_1)^N}{\sqrt{N!}} + c_2 \frac{ (\hat{a}^\dagger_1)^{N-1} \hat{a}^\dagger_2}{\sqrt{(N-1)!}} + \dots \right.  \\*  \left.+ c_{d_H} \frac{(\hat{a}^\dagger_M)^N}{\sqrt{N!}}\right] \ket{0,0,\dots,0_M} \quad \quad \nonumber
\end{eqnarray}
and applies the symbolic transformation, Eq.~(\ref{LO Creation Operator Transformation}). For large $N$ and $M$, however, this process can quickly become intractable. Here, we use an efficient and highly parallelizable numerical protocol for simulating linear optical quantum gates or state evolution that is equivalent to the action of Eq.~(\ref{LO Creation Operator Transformation}), but can effectively handle large values of $N$ and $M$. We will present this protocol in its entirety in Section~\ref{Section on Protocol}.

Before we begin, we define $A(U)$ as the unitary matrix which represents the action of Eq.~(\ref{LO Creation Operator Transformation}) on a quantum state in the Fock basis~(\ref{LO State Fock Basis}). The output state $\ket{\psi^\prime}$ of a linear optical circuit will be determined by standard matrix-vector multiplication,
\begin{equation}
\label{Matrix Vector Product}
\ket{\psi^\prime} = A(U) \ket{\psi}.
\end{equation}

If $N \ge 2$, the group $\textbf{A}=\{ A(U) \}$  of all possible linear optical operators is a proper (strict) subgroup of the unitary group,
\begin{equation}
\label{Proper Subgroup}
\textbf{A} \subset \textbf{U}(d_H).
\end{equation}
In other words, not all quantum transformations on multi-photon Fock states can be implemented via linear optical circuits. Indeed, the condition that entangling operations between photons be linearly accessible is  generally found to be severely restricting~\cite{Matt Smith,Braunstein,Jake Smith,Lougovski,Lutkenhaus,Carollo,Pavicic}. 
\section{Numerical Simulation of Linear Optical Devices}
\label{Section on Protocol}
For an optical circuit made up of $K$ components,
\begin{equation}
	\label{Fact 1}
A(U_K) A(U_{K-1}) \dots A(U_1) = A(U_1 U_2 \dots U_K).
\end{equation}
It is preferred to compile the optical hardware components via the matrix multiplication $U_1 U_2 \dots U_K$, rather than render each $A(U_k)$ independently. The proof of Eq.~(\ref{Fact 1}) appears in Appendix~\ref{Proof of Fact 1}.
\\ \\ \indent
Rewriting Eq.~(\ref{LO State Fock Basis}) as
\begin{equation}
\label{LO State Fock Vec Basis}
\ket{\psi} = \sum_{\vec{n}} c_{\vec{n}} \ket{\vec{n}} \,,
\end{equation}
where
\begin{equation}
\ket{\vec{n}} = \ket{n_1,n_2,\dots,n_M}
\end{equation}
are the Fock states, we define
\begin{equation}
\label{m definition}
\ket{\vec{m}(\vec{n})} = \ket{m_1,m_2,\dots,m_N}
\end{equation}
where $m_\alpha$ is the mode-location of photon number $\alpha$. Of course the choice of vector $\ket{\vec{m}(\vec{n})}$ for a given $\ket{\vec{n}}$ is not unique, since photons are indistinguishable and labeling them is an arbitrary process. We simply need to choose \textit{some} labeling and pick any valid $\ket{\vec{m}(\vec{n})}$ for each $\ket{\vec{n}}$. Then the elements of $A(U)$ in the basis $\outerproduct{\vec{n}^\prime}{\vec{n}}$ are given by
\begin{eqnarray}
\label{Fact 2}
A(U)_{\vec{n}^\prime,\vec{n}} = \quad \quad \quad \quad \quad \quad \quad \quad \quad \quad \quad \quad \quad \quad \quad \quad \quad \\ \small \nonumber \prod_{p=1}^{M} \frac{\sqrt{n_p^\prime !}}{\sqrt{n_p !}} \left[\sum_{\textrm{perm}(\vec{m}^\prime)} U_{m_1 m_1^\prime} U_{m_2 m_2^\prime} \dots U_{m_N m_N^\prime} \right] \,, 
\end{eqnarray}
where the summation is over all distinct permutations of integer entries in the vector $\vec{m}^\prime$. The proof of Eq.~(\ref{Fact 2}) appears in Appendix~\ref{Proof of Fact 2}.
\\
\\
\indent In practice, the input state $\ket{\psi}$ to an optical circuit  will often be a simple product state with a definite number of photons in each mode, e.g. $\ket{N,0,\dots,0_M}$. Furthermore, a partial measurement on the output state $\ket{\psi^\prime}$ will leave us in some relevant subspace of the full Hilbert space. We can then formally state the following facts.
\\
\\
\textit{Fact 1:} If the input states $\ket{\psi}$ to our optical circuit are known to be limited to some subspace of the full Hilbert space of the system, we need only build the relevant columns of $A(U)$.
\\
\\
\textit{Fact 2:} If the output states $\ket{\psi^\prime}$ of our optical circuit  are then projected onto some subspace of the full Hilbert space, we need only build the relevant rows of $A(U)$.
\\
\\ \indent
With Eq.~(\ref{Fact 1}), Eq.~(\ref{Fact 2}) and the \textit{Facts}, we can now present our protocol for simulating a linear optical quantum circuit.
\\
\\
\textbf{Initialization Stage:}
\\
\\ 
		(1)  Establish the Fock basis $\{\ket{\vec{n}} \}$ of our input state $\ket{\psi}$.  Include only the basis states having nonzero overlap with $\ket{\psi}$, as per \textit{Fact 1}.
		\newline
		\newline
		(2) For each element of the basis set $\ket{\vec{n}}$ construct a corresponding $\ket{\vec{m}(\vec{n})}$ as defined in Eq.~(\ref{m definition}).
		\newline
		\newline
		(3) Establish the Fock basis $\{\ket{\vec{n}^\prime}\}$ of our output state $\ket{\psi^\prime}$. Do not include basis states that will be projected out in the measurement, as per \textit{Fact 2}.
		\newline
		\newline
		(4)  For each element of the basis set $\ket{\vec{n}^\prime}$ construct a corresponding $\ket{\vec{m}^\prime(\vec{n}^\prime)}$ as defined in Eq.~(\ref{m definition}).
		\newline
		\newline
		(5) Store all $\vec{n}, \vec{n}^\prime, \vec{m}, \vec{m}^\prime$ as integer vectors.
		\newline
		\newline
\textbf{Rendering Stage:}
\newline
\newline
		(1)  Compile the total optical circuit composed of $K$ components by performing the matrix multiplication $ U = U_1 U_2 \dots U_K $.
		\newline
		\newline
		(2) Render the matrix representation of the quantum operator $A(U)$ using Eq.~(\ref{Fact 2}) for all of the relevant input basis states $\{\ket{\vec{n}}\}$ and all of the relevant output basis states $\{\ket{\vec{n}^\prime}\}$.
\newline
\newline \indent
 Initialization needs only to be performed once; we can then quickly build $A(U)$ for any particular optical circuit design, $U$. This allows fast, repeated simulation necessary for Monte Carlo applications or numerical optimization. For an open source implementation of this algorithm, see Ref.~\cite{GitHub}.
\section{Rail Encoding Qubits into Deterministic Blocks}
\label{Section Block Encoding}
It is well established that we can implement any quantum transformation acting on states in the Fock basis using linear optical components if only a single photon is in the system~\cite{Adami,Torma,Pittman}:
\begin{equation}
\label{Isomorphism}
	\mbox{If } N=1, \quad \textbf{A} = \textbf{U}(d_H)\,.
\end{equation}
That is, we can build a universal quantum computer or fully decodable quantum channel in a single-photon rail encoding. However, by forcing $N=1$, we sacrifice the scalability of our hardware; Eq.~(\ref{Hilbert Space Dimension}) simplifies to
\begin{equation}
d_H = M.
\end{equation}
To implement $j$ qubits, we need $M=2^j$ optical modes. Here, we propose grouping single-photon blocks of logical qubits together in order to establish scalability; we want to extend the applicability of single-photon computing to general, large-scale quantum algorithms. 

As an example, two qubits can be encoded into an $N=1,M=4$ system as shown in Table~\ref{Two Qubit Encoding Table}.
\begin {table}[h]
\begin{center}
	\begin{tabular}{l*{6}{c}r} 
		Logical Qubit State      \quad \quad \quad     & Physical Fock State \\
		\hline 
		\quad \quad \quad $\ket{00}$     & $\ket{1,0,0,0}$ \\
		\quad \quad \quad $\ket{01}$            & $\ket{0,1,0,0}$ \\
		\quad \quad \quad $\ket{10}$            & $\ket{0,0,1,0}$ \\
		\quad \quad \quad $\ket{11}$            & $\ket{0,0,0,1}$ \\
	\end{tabular}
	\caption{ \label{Two Qubit Encoding Table} Two qubits encoded into a single-photon block.}
\end{center}
\end{table}
We can then encode \textit{four} qubits into the $N=2,M=8$ system partitioned into two blocks as described in Table~\ref{Block Encoding Table}.
\begin {table}[h]
\begin{center}
	\begin{tabular}{l*{6}{c}r} 
		Logical Qubit State      \quad \quad \quad     & Physical Fock State \\
		\hline 
		\quad \quad \quad $\ket{0000}$     & $\ket{1,0,0,0,1,0,0,0}$ \\
		\quad \quad \quad $\ket{0001}$            & $\ket{1,0,0,0,0,1,0,0}$ \\
		\quad \quad \quad $\ket{0010}$           & 
		$\ket{1,0,0,0,0,0,1,0}$ \\
		\quad \quad \quad $\ket{0011}$           & 
		$\ket{1,0,0,0,0,0,0,1}$ \\
		\quad \quad \quad $\ket{0100}$           & 
		$\ket{0,1,0,0,1,0,0,0}$ \\
		\quad \quad \quad $\ket{0101}$           & 
		$\ket{0,1,0,0,0,1,0,0}$ \\
		\quad \quad \quad $\ket{0110}$           & 
		$\ket{0,1,0,0,0,0,1,0}$ \\
		\quad \quad \quad $\ket{0111}$            & $\ket{0,1,0,0,0,0,0,1}$ \\
		\quad \quad \quad $\ket{1000}$            & $\ket{0,0,1,0,1,0,0,0}$ \\
		\quad \quad \quad $\ket{1001}$            & $\ket{0,0,1,0,0,1,0,0}$ \\
		\quad \quad \quad $\ket{1010}$            & $\ket{0,0,1,0,0,0,1,0}$ \\
		\quad \quad \quad $\ket{1011}$            & $\ket{0,0,1,0,0,0,0,1}$ \\
		\quad \quad \quad $\ket{1100}$            & $\ket{0,0,0,1,1,0,0,0}$ \\
		\quad \quad \quad $\ket{1101}$            & $\ket{0,0,0,1,0,1,0,0}$ \\
		\quad \quad \quad $\ket{1110}$            & $\ket{0,0,0,1,0,0,1,0}$ \\
		\quad \quad \quad $\ket{1111}$            & $\ket{0,0,0,1,0,0,0,1}$ \\
	\end{tabular}
	\caption{ \label{Block Encoding Table} Four qubits encoded into two single-photon blocks.}
\end{center}
\end{table}
In this encoding, we can apply any single-qubit rotation or a controlled-Unitary gate between logical qubits 1 and 2 or between logical qubits 3 and 4 through deterministic manipulation of a single photon~\cite{Review Paper,Adami,Lanyon,Bao,Zadeh}. The difficulty here lies in the application of entangling operations between blocks. For example, we may want to apply the $\mbox{CNOT}_{1,4}$ gate meaning that the control is logical qubit 1 and the target is logical qubit 4. This operation requires the two photons in our system to interact; if a photon is contained in modes 3 or 4, we swap a photon between modes 5 and 6 or between modes 7 and 8. 

Generalizing the 2-qubit block encoding in Table~\ref{Two Qubit Encoding Table}, we can encode $q$ qubits into a block of 1 photon in $2^q$ modes. Again, operations between pairs of logical qubits contained entirely within in a block are simple to implement using linear optics, but we will need to test our ability to realize communication between blocks. We assume a  mapping between logical qubit states and Fock states within a single block as in Table~\ref{q Qubit Block Encoding}.
\begin {table}[h]
\begin{center}
	\begin{tabular}{l*{6}{c}r} 
		Logical Qubit State      \quad \quad \quad     & Physical Fock State \\
		\hline 
		\quad \quad \quad $\ket{0\dots00}$     & $\ket{1,0,0,0,\dots,0}$ \\
		\quad \quad \quad $\ket{0\dots01}$            & $\ket{0,1,0,0,\dots,0}$ \\
		\quad \quad \quad $\ket{0\dots10}$            & $\ket{0,0,1,0,\dots,0}$ \\
		\quad \quad \quad $\ket{0\dots11}$            & $\ket{0,0,0,1,\dots,0}$ \\
		\quad \quad \quad \quad \enspace \vdots & \vdots \\
		\quad \quad \quad $\ket{1\dots11}$            & $\ket{0,0,0,0,\dots,1}$ \\
	\end{tabular}
	\caption{ \label{q Qubit Block Encoding} q qubits encoded into a single-photon block.}
\end{center}
\end{table}
Without loss of generality, we can group two blocks together, choose the first qubit in the control block as a control qubit, and the last qubit in the target block as a target qubit. Then, we can generalize the $\mbox{CNOT}_{1,4}$ gate to the $\mbox{CNOT}_{\rm first,last}$ gate, which swaps adjacent modes in the target block if a photon is in the second half of the modes in the control block. This operation cannot be implemented through vanilla linear optics:
\begin{equation}
\mbox{CNOT}_{\rm first,last} \notin \textbf{A}.
\end{equation}
In the special case $q=1$, our block encoding reduces to the standard dual-rail encoded qubit~\cite{Review Paper}. In the KLM~\cite{KLM,KLM2} scheme, the $\mbox{CNOT}_{\rm first,last}$ gate can be applied to dual-rail qubits using probabilistic partial measurements with success probability $p=2/27$. 
\section{Elementary Photon-Entangling Operations}
\label{Section Elementary Photon Entangling Operations}

 In practice, we find it useful to dismantle the $\mbox{CNOT}_{\rm first,last}$ gate into a set of elementary entangling sub-operations. We can first examine the $\mathcal{C}_1$ sub-operation acting on three modes as defined in Table~\ref{One Control Two Targets}. 
\begin {table}[h]
\begin{center}
	\begin{tabular}{l*{6}{c}r} 
		$\ket{0,0,0}$  &  $\rightarrow$ & $\ket{0,0,0}$ \\ \\
		$\ket{0,0,1}$  & $\rightarrow$ & $\ket{0,0,1}$ \\
		$\ket{0,1,0}$ & $\rightarrow$ & $\ket{0,1,0}$ \\
		$\ket{1,0,0}$ & $\rightarrow$ & $\ket{1,0,0} $ \\ \\
		$\ket{1,0,1}$ & $\rightarrow$ & $\ket{1,1,0}$ \\
		$\ket{1,1,0}$ & $\rightarrow$ & $\ket{1,0,1}$ \\
	\end{tabular}
	\caption{ \label{One Control Two Targets} The $\mathcal{C}_1$  operation acting on three modes expressed as a transformation of Fock basis states. The first mode acts as the control while the second and third modes act as targets. We can build the $\mbox{CNOT}_{\rm first,last}$ gate by applying this operation $2^{2 q -2}$ times for block size $q \ge 1 $.}
\end{center}
\end{table}
Here, the first mode is a control mode, while the second and third modes are target modes. We apply the $\mathcal{C}_1$  operator four times to build the $\mbox{CNOT}_{1,4}$ in $q=2$, where in each sub-operation the control mode take values 3 or 4 in Table~\ref{Block Encoding Table} and the pair of target modes takes values 5,6 or 7,8. The end result is the $\mbox{CNOT}_{1,4}$ operation: If a photon is in mode 3 or 4 we swap a photon between mode 5 and mode 6 or a photon between mode 7 and mode 8. More generally, a $\mbox{CNOT}_{\rm first,last}$ gate for block size $q$ can be constructed by applying the $\mathcal{C}_1$  operation $2^{2q-2}$ times by setting the control mode to each of the modes in the second half of the control block, and the two target modes to every sequential pair of modes in the target block. Because there will never be more than a single photon in a block, the basis states in Table~\ref{One Control Two Targets} will be the only possible inputs to the $\mathcal{C}_1$  transformation. If $\mathcal{C}_1$ is performed correctly, photon leakage between blocks cannot occur. Other entangling sub-operations we study here include the $\mathcal{C}_2$, $\mathcal{C}_3$, and $\mathcal{C}_4$ gates which are presented in Tables~\ref{Two Controls Two Targets}-\ref{Two Controls Four Targets}. As the sub-operations get increasingly more complex, a smaller number of such sub-operations is required to implement the inter-block logical operation  $\mbox{CNOT}_{\rm first,last}$.
\begin {table}[h]
\begin{center}
	\begin{tabular}{l*{6}{c}r} 
		$\ket{0,0,0,0}$  &  $\rightarrow$ & $\ket{0,0,0,0}$ \\ \\
		$\ket{0,0,0,1}$  & $\rightarrow$ & $\ket{0,0,0,1}$ \\
		$\ket{0,0,1,0}$ & $\rightarrow$ & $\ket{0,0,1,0}$ \\
		$\ket{0,1,0,0}$ & $\rightarrow$ & $\ket{0,1,0,0} $ \\ 
		$\ket{1,0,0,0}$ & $\rightarrow$ & $\ket{1,0,0,0} $ \\ \\
		$\ket{1,0,0,1}$  & $\rightarrow$ & $\ket{1,0,1,0}$ \\
		$\ket{1,0,1,0}$ & $\rightarrow$ & $\ket{1,0,0,1}$ \\
		$\ket{0,1,0,1}$ & $\rightarrow$ & $\ket{0,1,1,0} $ \\ 
		$\ket{0,1,1,0}$ & $\rightarrow$ & $\ket{0,1,0,1} $ \\ \\
	\end{tabular}
	\caption{ \label{Two Controls Two Targets} The $\mathcal{C}_2$  operation acting on four modes expressed as a transformation of Fock basis states. The first two modes act as the controls while the last two modes act as targets. We can build the $\mbox{CNOT}_{\rm first,last}$ gate by applying this operation $2^{2 q - 3}$ times for block size $ q \ge 2 $.}
\end{center}
\end{table}
\begin {table}[h]
\begin{center}
	\begin{tabular}{l*{6}{c}r} 
		$\ket{0,0,0,0,0}$  &  $\rightarrow$ & $\ket{0,0,0,0,0}$ \\ \\
		$\ket{0,0,0,0,1}$  & $\rightarrow$ & $\ket{0,0,0,0,1}$ \\
		$\ket{0,0,0,1,0}$ & $\rightarrow$ & $\ket{0,0,0,1,0}$ \\
		$\ket{0,0,1,0,0}$ & $\rightarrow$ & $\ket{0,0,1,0,0} $ \\ 
		$\ket{0,1,0,0,0}$ & $\rightarrow$ & $\ket{0,1,0,0,0} $\\
		$\ket{1,0,0,0,0}$ & $\rightarrow$ & $\ket{1,0,0,0,0} $ \\ \\
		$\ket{1,0,0,0,1}$  & $\rightarrow$ & $\ket{1,0,0,1,0}$ \\
		$\ket{1,0,0,1,0}$  & $\rightarrow$ & $\ket{1,0,0,0,1}$ \\
		$\ket{1,0,1,0,0}$  & $\rightarrow$ & $\ket{1,1,0,0,0}$ \\
		$\ket{1,1,0,0,0}$  & $\rightarrow$ & $\ket{1,0,1,0,0}$ \\
	\end{tabular}
	\caption{ \label{One Control Four Targets} The $\mathcal{C}_3$ operation acting on five modes expressed as a transformation of Fock basis states. The first mode acts as the control while the last four modes act as targets. We can build the $\mbox{CNOT}_{\rm first,last}$ gate by applying this operation $2^{2 q - 3}$ times for block size $q \ge 2$.}
\end{center}
\end{table}
\begin {table}[h]
\begin{center}
	\begin{tabular}{l*{6}{c}r} 
		$\ket{0,0,0,0,0,0}$  &  $\rightarrow$ & $\ket{0,0,0,0,0,0}$ \\ \\
		$\ket{0,0,0,0,0,1}$  & $\rightarrow$ & $\ket{0,0,0,0,0,1}$ \\
		$\ket{0,0,0,0,1,0}$ & $\rightarrow$ & $\ket{0,0,0,0,1,0}$ \\
		$\ket{0,0,0,1,0,0}$ & $\rightarrow$ & $\ket{0,0,0,1,0,0} $ \\ 
		$\ket{0,0,1,0,0,0}$ & $\rightarrow$ & $\ket{0,0,1,0,0,0} $\\
		$\ket{0,1,0,0,0,0}$ & $\rightarrow$ & $\ket{0,1,0,0,0,0} $ \\
		$\ket{1,0,0,0,0,0}$ & $\rightarrow$ & $\ket{1,0,0,0,0,0} $ \\ \\
		$\ket{0,1,0,0,0,1}$  & $\rightarrow$ & $\ket{0,1,0,0,1,0}$ \\
		$\ket{0,1,0,0,1,0}$  & $\rightarrow$ & $\ket{0,1,0,0,0,1}$ \\
		$\ket{0,1,0,1,0,0}$  & $\rightarrow$ & $\ket{0,1,1,0,0,0}$ \\
		$\ket{0,1,1,0,0,0}$  & $\rightarrow$ & $\ket{0,1,0,1,0,0}$ \\
		$\ket{1,0,0,0,0,1}$  & $\rightarrow$ & $\ket{1,0,0,0,1,0}$ \\
		$\ket{1,0,0,0,1,0}$  & $\rightarrow$ & $\ket{1,0,0,0,0,1}$ \\
		$\ket{1,0,0,1,0,0}$  & $\rightarrow$ & $\ket{1,0,1,0,0,0}$ \\
		$\ket{1,0,1,0,0,0}$  & $\rightarrow$ & $\ket{1,0,0,1,0,0}$ \\
	\end{tabular}
	\caption{ \label{Two Controls Four Targets} The $\mathcal{C}_4$ operation acting on six modes expressed as a transformation of Fock basis states. The first two modes act as the controls while the last four modes act as targets. We can build the $\mbox{CNOT}_{\rm first,last}$ gate by applying this operation $2^{2 q - 4}$ times for block size $q \ge 2$.}
\end{center}
\end{table}

\begin{figure}[h]
 	\centering
 	\includegraphics[width=0.5 \textwidth]{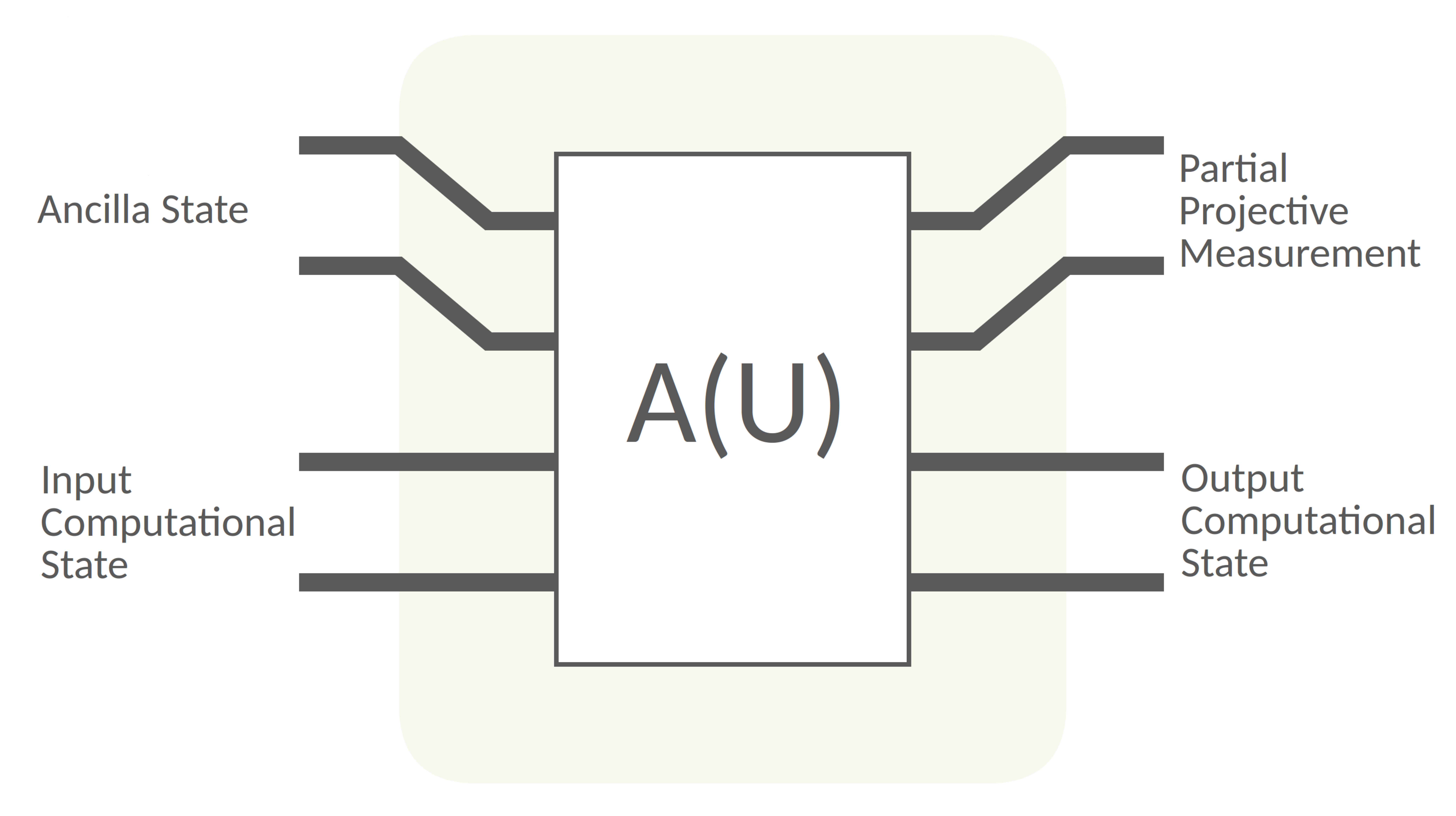}
 	\caption{The measurement-assisted transformation described by $E$. An input computational state $\ket{\psi_c}$ and ancilla state $\ket{\psi_a}$ are processed through a linear circuit. Then, a partial measurement $P$ is performed and we are left with the output computational state $\ket{\psi_c^\prime}$.}
 	\label{Figure - PAULA Operator}
 \end{figure}

The focus of our work here is to test our ability to build these operators as probabilistic measurement-assisted transformations. By this, we mean that we augment our optical computational state $\ket{\psi_c}$ with an ancilla state $\ket{\psi_a}$, apply a linear optical transformation, and then perform a partial projective measurement (refer to Fig.~\ref{Figure - PAULA Operator}). 
The total action of this non-unitary process can be written formally as a Kraus operator we call \textit{PAULA}, which acts on any input optical quantum state $\ket{\psi_{c}}$ composed of $N_c$ photons in $M_c$ modes contained in the computational subspace:
\begin{equation}
\label{Kraus Paula Operator}
E = P A(U) L_a
\end{equation} 
\begin{equation}
\ket{\psi_{c}^\prime} = \frac{E \ket{\psi_{c}}}{\sqrt{ \bra{\psi_c} E^\dagger E \ket{\psi_c} }}.
\end{equation}
In Eq.~(\ref{Kraus Paula Operator}), $L_a$ is the operator defined by
\begin{equation}
\label{Ancilla Linker}
L_a \ket{\psi_{c}} = \ket{\psi_a} \otimes \ket{\psi_{c}} \quad \quad  \forall \, \ket{\psi_{c}} \,,
\end{equation}
where $\ket{\psi_a}$ is a normalized ancilla state composed of $N_a$ photons in $M_a$ optical modes. The matrix representation of $L_a$ is sparse and satisfies $L_a^\dagger L_a = I$.
$A(U)$ is the standard unitary linear optical transformation as defined in Eq.~(\ref{Fact 2}), where $N=N_c+N_a$ and $M=M_c+M_a$. $P$ is a partial projective measurement over the ancilla modes. We assume the projection operator $P$ is given by
\begin{equation}
P = \outerproduct{\vec{n}_a}{\vec{n}_a} \otimes I\,,
\end{equation}
where $\ket{\vec{n}_a}$ is some Fock state 
containing the same number of photons $N_a$ and modes $M_a$ as the input ancilla state $\ket{\psi_a}$. $E$ therefore preserves photon and mode number. In accordance with \textit{Facts 1} and \textit{2}, we do not construct the entire matrix $A(U)$ -- only the columns associated with nonzero components of $L_a \ket{\psi_{c}}$ and the rows left after the projection $P$. We define $d_c$ as the dimension of the subspace containing all possible input computational states to our circuit, and $d_c^\prime$ as the full physical Hilbert space associated with the output modes:
\begin{equation}
	d_c^\prime = \frac{(N_c+M_c-1)!}{N_c! (M_c-1)!} \,.
\end{equation}
$E$ can be represented as a $d_c^\prime$ by $d_c$ rectangular matrix.

A measure of gate fidelity (more precisely, the real part of the fidelity amplitude) between the \textit{PAULA} operator $E$ and a target operation $T$ can be defined by
\begin{equation}
\label{Gate Fidelity}
F(E,T) = \frac{\Re[\tr(E^\dagger T)]}{\sqrt{d_c \tr(E^\dagger E)} }.
\end{equation}
The probability of having successfully applied $E$ for $F(E,T) \rightarrow 1$ is given by
\begin{equation}
\label{Success Probability} 
S(E) = \frac{\tr(E^\dagger E)}{d_c}.
\end{equation}

To construct the $\mathcal{C}_1$ operation defined in Table~\ref{One Control Two Targets}, which acts on input states having 0, 1, or 2 photons, we simultaneously maximize the gate fidelity and success probability for each  $N_c=0,1,2$ \textit{separately} over the same quantum optical circuit $U$ and ancilla state $\ket{\psi_a}$. That is, we establish an ideal target $T_{N_c}$ and Kraus \textit{PAULA} operator $E_{N_c}$ for each photon number $N_c$ and attempt to find an optical mode transformation $U$ and ancilla input $\ket{\psi_a}$ such that  $F(E_{N_c},T_{N_c}) \rightarrow 1$ for all $N_c$ with optimal $S(E_{N_c})$. We will demonstrate the details of this in the following section for $\mathcal{C}_1$. The same approach is readily extended to $\mathcal{C}_2 \dots \mathcal{C}_4$, and in the following we omit a detailed discussion of the optimization relating to those operators, instead showing only the final results.
\section{Numerical Analysis of $\mathcal{C}_1$}
\label{Section Numerical Testing}
We consider first $\mathcal{C}_1$ acting on the subspace of $N_c=0$ photons. Here we get the correct transformation for free:
\begin{equation}
\label{Free Operation}
\ket{0,0,0} \rightarrow \ket{0,0,0}.
\end{equation}
Up to a possible overall phase (see below), this will be the result for any linear optical circuit  $U$ and any ancilla state $\ket{\psi_a}$, just because $E$ in this subspace is always a 1 by 1 matrix: $d_c = d_c^\prime = 1$.

If only a single photon is found in the three modes, $N_c=1$, we have
\begin{equation}
d_c = d_c^\prime = 3\,,
\end{equation}
and we strive for a target operator,
\begin{equation}
\label{T1 1C2T}
T_1=	\begin{pmatrix} 1 & 0 & 0  \\ 0 & 1 & 0  \\ 0 & 0 & 1   \end{pmatrix}  
\end{equation}
in the basis
\begin{equation}
\{ \ket{\vec{n}_c} \} = \{ \ket{\vec{n}_c^\prime} \} = \{ \ket{0,0,1},\ket{0,1,0},\ket{1,0,0} \}.
\end{equation}
Finally, in the two-photon subspace, $N_c=2$, we have
\begin{equation}
d_c = 2 \quad d_c^\prime = 6 \,,
\end{equation}
and strive for a target operator,
\begin{equation}
\label{T2 1C2T}
T_2=\begin{pmatrix} 0 & 1  \\ 1 & 0  \\ 0 & 0 \\ 0 & 0 \\ 0 & 0 \\ 0 & 0   \end{pmatrix}  
\end{equation}
in the basis
\begin{equation}
\{ \ket{\vec{n}_c} \} = \{ \ket{1,0,1},\ket{1,1,0} \}
\end{equation}
\begin{eqnarray}
& \{ \ket{\vec{n}^\prime_c } \} = \{ \ket{1,0,1},\ket{1,1,0}, \\ &\ket{2,0,0},\ket{0,2,0},\ket{0,1,1},\ket{0,0,2} \}. \nonumber
\end{eqnarray}

We define the \textit{PAULA} operators $E_{N_c}(U,\psi_a)$ separately for the $N_c=0$, $N_c=1$ and $N_c=2$ subspaces,
and numerically maximize the function
\begin{equation}
\label{Merit Function For Blocks}
\small f(U,\psi_a) = \sum_{N_c=0}^{2} \Big[ F(E_{N_c},T_{N_c}) + \epsilon S(E_{N_c}) \Big]
\end{equation}
where $\epsilon > 0$ is a real-valued numerical weight for the optimization, or equivalently a Lagrange multiplier. We systematically vary parameters $\epsilon$, $N_a$, $M_a$, and partial measurement $P$, and repeat maximization of Eq.~(\ref{Merit Function For Blocks}) to find global solutions.

Any difference in global phases between the operations in Eqs.~(\ref{Free Operation}), (\ref{T1 1C2T}), and (\ref{T2 1C2T}) will cause an unwanted relative phase shift between Fock basis states in the full transformation $\mathcal{C}_1$. For this reason, we use the gate fidelity amplitude defined in Eq.~(\ref{Gate Fidelity}), which accounts for a global phase difference between $E$ and $T$. 
\section{Results}
\label{Section on Results}
\subsection{The entangling operation $\mathcal{C}_1$}
The physical operation $\mathcal{C}_1$ acting on three modes is of particular importance; it can be applied once to two qubits encoded in the dual rail ($q=1$) in order to apply an entangling $\mbox{CNOT}$ gate between the two qubits. Numerical optimization of Eq.~(\ref{Merit Function For Blocks}) leads to the same optimal solution as the one given in~\cite{Uskov}, with two ancilla photons in two modes:
\begin{eqnarray}
& F(E_{N_c},T_{N_c})_{\rm max} = 1 \nonumber \\
& S(E_{N_c})_{\rm max} = 2/27 \\
& N_a=M_a=2 \nonumber \\
& P = \outerproduct{11}{11} \nonumber	.
\end{eqnarray}
Increasing the number of ancilla resources beyond $N_a=M_a=2$ does not improve the success probability for numerically accessible ancilla sizes ($N_a,M_a \le 8$). Thus, for a block size of $q \ge 1$ qubits, we find a maximum probability of successfully applying a $\mbox{CNOT}_{\rm first,last}$ gate composed of $\mathcal{C}_1$ operations using measurement-assisted transformations to be
\begin{equation}
\label{1C2T Result}
p = (2/27)^{2^{2q-2}}.
\end{equation}
\subsection{The entangling operation $\mathcal{C}_2$}
We find the optimal solution:
\begin{eqnarray}
& F(E_{N_c},T_{N_c})_{\rm max} = 1 \nonumber \\
& S(E_{N_c})_{\rm max} = 0.0221391 \\
&N_a = 3,M_a=4 \nonumber \\
&P = \outerproduct{1110}{1110} \nonumber	.
\end{eqnarray}
For a block size of $q \ge 2$ qubits, we find a maximum probability of successfully applying a $\mbox{CNOT}_{\rm first,last}$ gate composed of $\mathcal{C}_2$ operations  using measurement-assisted transformations to be
\begin{equation}
\label{2C2T Result}
p = (0.0221391)^{2^{2q-3}}.
\end{equation}
\subsection{The entangling operation $\mathcal{C}_3$}
We find the optimal solution:
\begin{eqnarray}
& F(E_{N_c},T_{N_c})_{\rm max} = 1 \nonumber \\
& S(E_{N_c})_{\rm max} = 0.0221266 \\
& N_a = 3,M_a=4 \nonumber \\
& P = \outerproduct{1110}{1110} .\nonumber	
\end{eqnarray}
For a block size of $q \ge 2 $ qubits, we find a maximum probability of successfully applying a $\mbox{CNOT}_{\rm first,last}$ gate composed of $\mathcal{C}_3$ operations  using measurement-assisted transformations to be
\begin{equation}
\label{1C4T Result}
p = (0.0221266)^{2^{2q-3}}.
\end{equation}
\subsection{The entangling operation $\mathcal{C}_4$}
We find the optimal solution:
 \begin{eqnarray}
 & F(E_{N_c},T_{N_c})_{\rm max} = 1 \nonumber\\
 & S(E_{N_c})_{\rm max} = 0.00691511 \\
 & N_a = M_a=4 \nonumber\\
 & P = \outerproduct{1111}{1111}. \nonumber	
 \end{eqnarray}
For a block size of $q \ge 2$ qubits, we find a maximum probability of successfully applying a $\mbox{CNOT}_{\rm first,last}$ gate composed of $\mathcal{C}_4$ operations  using measurement-assisted transformations to be
 \begin{equation}
 \label{2C4T Result}
 p = (0.00691511)^{2^{2q-4}}.
 \end{equation}
 \section{Discussion}
 \label{Section Conclusion}
\begin{figure}[h]
  	\centering
  	\includegraphics[width=0.5 \textwidth]{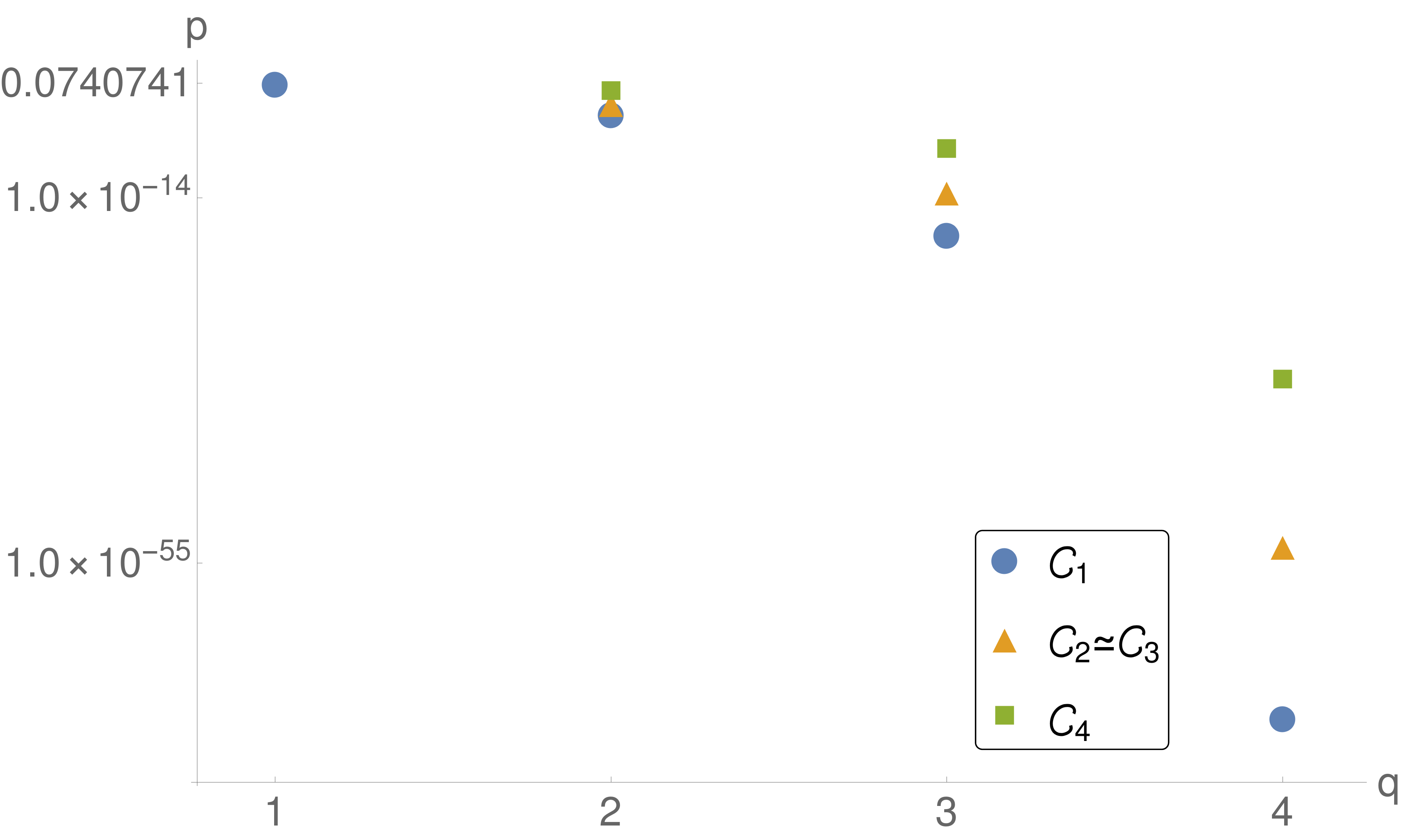}
  	\caption{Maximized success probabilities $p$ for implementing a full $\mbox{CNOT}_{\rm first,last}$ gate composed entirely of the sub-operations $\mathcal{C}_1$, $\mathcal{C}_2$, $\mathcal{C}_3$, $\mathcal{C}_4$, for blocks encoding $q$ qubits.}
  	\label{Figure - Block Encoding Results}
  \end{figure}

\begin{figure}[ht]
 	\centering
 	\includegraphics[width=0.25 \textwidth]{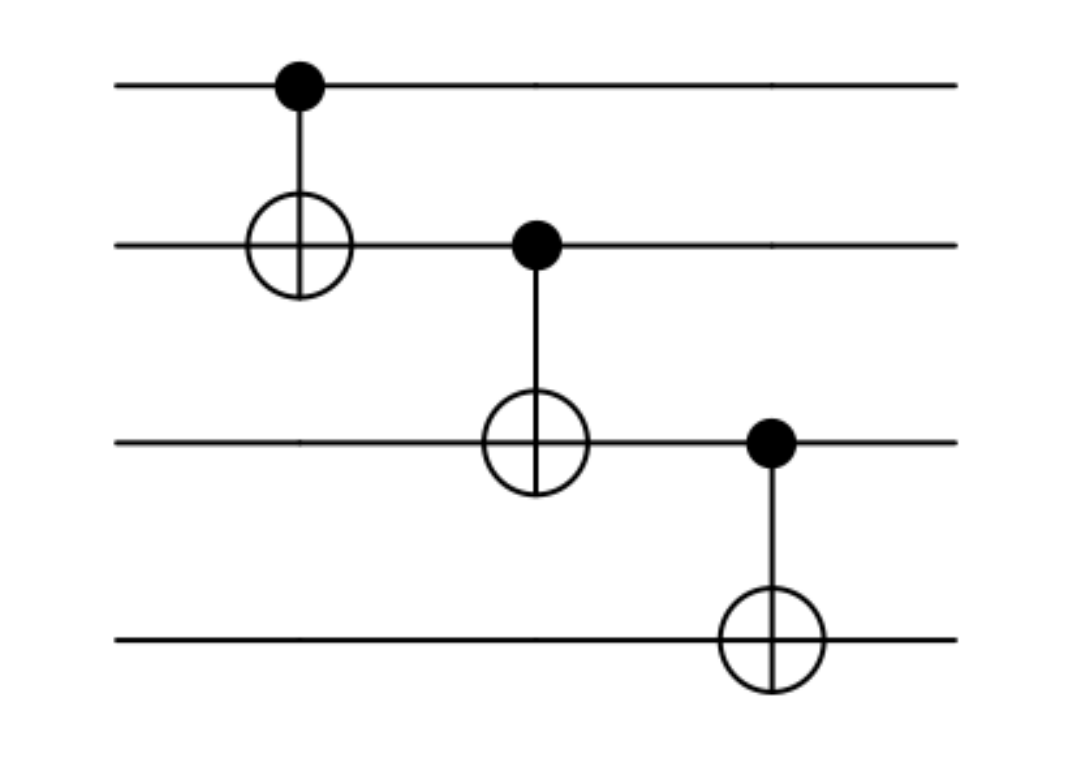}
 	\caption{The quantum circuit $\mbox{CNOT}^3$ acting on four qubits.}
 	\label{Three CNOTs}
 \end{figure}

 We observe a promising increase in success probability of applying the $\mbox{CNOT}_{\rm first,last}$ gate as we move to sub-operations acting on a greater number of modes; refer to Fig.~\ref{Figure - Block Encoding Results}. 
Even within the reach of our numerical simulations, we find an improvement over standard KLM for applying entangling operations within the context of some quantum algorithms. For example, the simple quantum circuit 
 \begin{eqnarray}
 \small & \mbox{CNOT}^3 = \\ \small \nonumber & (I \otimes I \otimes \mbox{CNOT}) \cdot (I \otimes  \mbox{CNOT} \otimes I) \cdot (\mbox{CNOT} \otimes I \otimes I)
 \end{eqnarray}
 acting on four qubits is presented in Fig.~\ref{Three CNOTs}. Through the standard KLM protocol acting on qubits encoded in the dual-rail, each CNOT gate can be implemented with success probabilty $2/27$ using ancilla resources $N_a=M_a=2$. The total success probability of the $\mbox{CNOT}^3$ gate using KLM is then $(2/27)^3$ using ancilla resources $ N_a=M_a = 6 $. Grouping these four qubits into two deterministic blocks of two qubits, $q=2$, we can apply the same operation with a success probability of $0.00691511$ using ancilla resources $ N_a=M_a=4 $. This is almost a twenty-fold improvement, while using a smaller ancilla resource.

 The trend in Fig.~\ref{Figure - Block Encoding Results} suggests an order of magnitude improvement in success probability as we increase the size (i.e., number of modes operated on) of entangling sub-operations between blocks. Future work can further increase the simulation size to confirm this trend. A high probability $\mbox{CNOT}_{\rm first,last}$ gate built from larger sub-operations for block size $q \ge 3$ would have significant implications for linear optical quantum computing.
\acknowledgments
We are thankful for helpful discussions with S. Kaminsky, E. Knutson, and J. Shipman. This research was supported in part using high performance computing (HPC) resources and services provided by Technology Services at Tulane University, New Orleans, LA.
\appendix
\section{Proof of Eq.~(\ref{Fact 1})}
\label{Proof of Fact 1}
Eq.~(\ref{Fact 1}) can be proved by induction. The base case is trivial; for a lone unitary matrix $U_1$, Eq.~(\ref{Fact 1}) reads
\begin{equation}
	A(U_1) = A(U_1).
\end{equation}
We can then assume
\begin{equation}
\label{Inductive Step}
	A(U_{K-1}) A(U_{K-2}) \dots A(U_1) = A(U_1 U_2 \dots U_{K-1}).
\end{equation}
That is, applying the left hand side of~(\ref{Inductive Step}) is equivalent to the total transformation
\begin{equation}
	\hat{a}^\dagger_\alpha \rightarrow \sum_\beta (U_1 U_2 \dots U_{K-1})_{\alpha,\beta} \enspace \hat{a}^\dagger_\beta.
\end{equation}
Now we add one final optical component and apply the transformation $A(U_K)$. Our total transformation is now
\begin{equation}
	\hat{a}^\dagger_\alpha \rightarrow \sum_\beta (U_1 U_2 \dots U_{K-1})_{\alpha,\beta} \Big(\sum_\gamma U_{\beta,\gamma}^K \hat{a}^\dagger_\gamma \Big)\,,
\end{equation}
where $U_{\beta,\gamma}^K$ are the $\beta,\gamma$ elements of $U_K$. Then,
\begin{eqnarray}
\hat{a}^\dagger_\alpha \rightarrow \sum_{\beta,\gamma} (U_1 U_2 \dots U_{K-1})_{\alpha,\beta} U^K_{\beta,\gamma} \hat{a}^\dagger_\gamma
\\*
= \sum_\gamma (U_1 U_2 \dots U_K)_{\alpha,\gamma} \hat{a}^\dagger_\gamma \,, \quad \quad
\end{eqnarray}
which is equivalent to $A(U_1 U_2 \dots U_K). \quad \quad\blacksquare $
\section{Proof of Eq.~(\ref{Fact 2})}
\label{Proof of Fact 2}
We can write Eq.~(\ref{LO State Fock Vec Basis}) in the $\ket{\vec{m}(\vec{n})}$ basis:
\begin{equation}
\ket{\psi} = \sum_{\vec{n}} c_{\vec{n}} \ket{\vec{m}(\vec{n})} \,.
\end{equation}
We note that because of photon indistinguishability, the mapping from $\ket{\vec{n}}$ to $\ket{\vec{m}(\vec{n})}$ is not unique.  This mapping is, however, injective. No matter which of the allowed mappings we choose, no information about the state $\ket{\psi}$ is lost. Thus, we are free to pick any mapping we want for each basis Fock state. Then
\begin{equation}
\ket{\psi} = \sum_{\vec{n}} \frac{\vec{c}_n}{\prod_{p=1}^{M} \sqrt{n_p !}} \hat{a}_{m_1}^\dagger \hat{a}_{m_2}^\dagger \dots \hat{a}_{m_N}^\dagger |\vec{0} \rangle \,.
\end{equation}
We define the operators
\begin{equation}
\hat{U}_\alpha = U_{\alpha 1} \hat{a}^\dagger_1 + U_{\alpha 2} \hat{a}^\dagger_2 \dots U_{\alpha M} \hat{a}^\dagger_M.
\end{equation}
Then
\begin{eqnarray}
\label{Fact 2 Proof Checkpoint}
& \ket{\psi^\prime} = A(U) \ket{\psi} =  \quad \quad \quad \quad \quad \quad \quad \quad \quad \quad \\ & \nonumber
\sum_{\vec{n}} \frac{c_{\vec{n}}}{\prod_{p=1}^{M} \sqrt{n_p !}} \enspace \hat{U}_{m_1}  \hat{U}_{m_2}  \dots  \hat{U}_{m_N} |\vec{0} \rangle.
\end{eqnarray}
The product of operators $\hat{U}_\alpha$ in~(\ref{Fact 2 Proof Checkpoint}) will return a massive expression of $M^N$ terms. We can use the commutativity of the creation operators 
\begin{equation}
[\hat{a}_i ^ \dagger,\hat{a}_j ^ \dagger] = 0 \quad \quad \forall\, i,j \,
\end{equation}
to compress it:
\begin{eqnarray}
\ket{\psi^\prime} = \sum_{\vec{n}} \frac{c_{\vec{n}}}{\prod_{p=1}^{M} \sqrt{n_p !}} \cdot \quad \quad \quad \quad \quad \quad \quad \quad \quad \\  \nonumber \sum_{\substack{1 \le m_1^\prime \le \cdots \\ \le m_N^\prime \le M}}
\left[ \sum_{\textrm{perm}(\vec{m}^\prime)} U_{m_1 m_1^\prime} \dots U_{m_N m_N^\prime} \right] \hat{a}_{m_1^\prime}^\dagger \dots \hat{a}_{m_N^\prime}^\dagger |\vec{0} \rangle.
\end{eqnarray}
Finally, we use 
\begin{equation}
\hat{a}_{m_1^\prime}^\dagger \dots \hat{a}_{m_N^\prime}^\dagger |\vec{0} \rangle =  \prod_{p=1}^{M} \sqrt{ n_p^\prime(\vec{m}^\prime)! } \ket{\vec{n}^\prime(\vec{m}^\prime)} 
\end{equation}
 to obtain
\begin{eqnarray}
\label{Fact 2 Proof Checkpoint 2}
\small \ket{\psi^\prime} &=& \sum_{\vec{n}} c_{\vec{n}}   \sum_{1 \le m_1^\prime \le \cdots \le m_N^\prime \le M}
\left[ \prod_{p=1}^{M} \frac{\sqrt{n_p^\prime(\vec{m}^\prime) !}}{ \sqrt{n_p !}} \right] \\  \nonumber &\cdot&  \left[ \sum_{\textrm{perm}(\vec{m}^\prime)} U_{m_1 m_1^\prime} \dots U_{m_N m_N^\prime} \right] 
 \ket{\vec{n}^\prime (\vec{m}^\prime) } .
\end{eqnarray}
We recognize Eq.~(\ref{Fact 2 Proof Checkpoint 2}) as the result of a matrix-vector product, Eq.~(\ref{Matrix Vector Product}), if we define $A(U)$ as in Eq.~(\ref{Fact 2}).  $ \blacksquare $

\end{document}